\documentclass[a4paper]{jpconf}

\usepackage{amsmath}
\usepackage{graphics}
\begin{document}
\title{A causal net approach to relativistic quantum mechanics}
\author{R D Bateson
\footnote[1]{Present address:
Oxford Man Institute, University of Oxford, Eagle House, Walton Well Road, OX2 6ED, UK.}}
\address{London Centre of Nanotechnology, University College London, 17-19 Gordon St., London, WC1H 0AH, UK.}

\ead{richarddbateson@gmail.com}

\begin{abstract}

In this paper we discuss a causal network approach to describing relativistic quantum mechanics. Each vertex on the causal net represents a possible point event or particle observation. By constructing the simplest causal net based on Reichenbach-like conjunctive forks in proper time we can exactly derive the 1+1 dimension Dirac equation for a relativistic fermion and correctly model quantum mechanical statistics. Symmetries of the net provide various quantum mechanical effects such as quantum uncertainty and wavefunction, phase, spin, negative energy states and the effect of a potential. The causal net can be embedded in 3+1 dimensions and is consistent with the conventional Dirac equation. In the low velocity limit the causal net approximates to the Schr\"{o}dinger equation and Pauli equation for an electromagnetic field. Extending to different momentum states the net is compatible with the Feynman path integral approach to quantum mechanics that allows calculation of well known quantum phenomena such as diffraction.

\end{abstract}

\section{Introduction}

Causality, or the concept of ``cause and effect'', has long been viewed by philosophers as a fundamental and often an {\emph{a priori}} principle in our understanding and interpretation of nature. Although Newton's laws were clearly written in causal terms, the indeterminacy of quantum mechanics lead to confusion of the role of causality in describing quantum systems. Several attempts have been made to fuse relativity and quantum indeterminism, for example, where branching space-time can be built on the primitives of a set of ``possible point events'' and causal relations [1] and the recent causal set approach to quantum gravity [2]. The fundamental equation of relativistic quantum mechanics is the Dirac equation for a fermion [3] and is based on the concept of continuous space and time. Richard Feynman [4] presented a discrete space-time derivation of the 1+1 dimension Dirac equation for a free particle -- the ``Feynman chequerboard'' -- since a luminal velocity massive particle is viewed in the calculation as ``zig-zagging'' diagonally forwards through space-time in a similar manner to a bishop in chess. Numerous attempts have been made to achieve a discrete quantum mechanics [5,6,7,8,9] but an exact lattice based formulation has never been achieved.  In this paper we discuss a causal network discretisation approach which exactly derives the full 4-vector Dirac equation and provides all the common fermion features, such as spin, negative energy states, action of a potential and summation of paths. The most basic causal net describes a plane wave solution with the space axis aligned along the direction of momentum. This 1+1 dimension net can be embedded in 3+1 dimension space-time using the Pauli matrices and is consistent with the full Dirac equation and quantum mechanical statistics.

\section{Reichenbach's principle of common cause and causal networks}
The application of probability theory to causality and its relation to the direction of time was developed by Hans Reichenbach. His \emph{principle of common cause} (PCC) [10] was summarized as follows: ``If coincidences of the two events A and B occur more frequently than would correspond to their independent occurrence, that is, if these events satisfy $P(A.B) > P(A)P(B)$ then there exists a common cause C for these events that the fork ACB is conjunctive.'' That is the probability of A and B occurring together is greater than the product of the individual probabilities of A and B. A conjunctive fork ACB (see Fig. 1) between events is open on one side where C is earlier in time than A or B. This asymmetry Reichenbach argued provides a definition of the flow of the direction of time in terms of microstatistics. Essentially a common cause is expected when coincidences or correlations between events occur repeatedly with greater frequency than complete statistical independence $P(A.B) = P(A)P(B)$. The principle of common cause provides a definition of simultaneity, since if A and B are simultaneous there cannot be a causal linkage between them except through the earlier event C. Reichenbach's principle of common cause can be readily extended into a relativistic framework [11] and a standard simultaneity condition developed [12] where simultaneous events lie on a hyperplane orthogonal to the particle world-line. 
It is widely accepted that quantum mechanics cannot be developed from a basic application of Reichenbach's principle of common cause with a single conjunctive fork since general quantum mechanical statistics violate the principle. The principle of common cause involving a single conjunctive fork can even actually be used to formulate the well known Bell's theorem [13,14], which motivated the famous experiments by Alain Aspect [15] to exclude the possibilities of certain types of hidden variables. Here we shall consider a modified framework where a complete causal network of possible events is comprised of conjunctive forks such that each possible event has two effective local common causes or screening factors. It appears that, at least in the simple case we consider, this allows our common cause principle –- based on the simultaneity of neighbouring possible events –- to be applied consistently with quantum statistics.

We shall adopt a relational view of time as an ordered series of closely spaced ``events''. Now if we consider time as a series of closely spaced events then from this perspective a classical particle trajectory could appear as a statistically correlated series of events in space-time (for example, a series of actual observations). If the correlation is perfect then one may loosely say that an event at one point in space ``causes'' the event at the next point, providing a Newtonian trajectory. However, if the correlation of events is imperfect, but greater than that resulting from statistical independence, then adjacent events in space are implied to have a common cause originating at a previous time. A trajectory becomes probabilistic in nature and we would have to involve a statistical interpretation. A network of Reichenbach's conjunctive forks [10] constitutes a causal net in which time is ordered and events may be considered simultaneous only when they share a common cause. To construct the causal net for a particle motion in space-time, we consider a 1 dimensional space aligned with the direction of particle motion, and embedded in 3 dimensional space. In this 1 dimensional space the simplest causal net that satisfies our definition of simultaneity is a 1+1 dimensional ``diamond'' lattice with causal links connecting the lattice points as in Figure 2. Each causal connection is defined by a connecting arrow giving a definite lineal order and an associated probability. Each vertex on the causal net represents a possible event –- meaning a possible observation of the particle –- and has two incoming and two outgoing causal connections so that each event has two effective possible common causes. Starting at a vertex and following an outgoing arrow at random at each subsequent vertex describes a ``causal chain'' as a series of possible events.  Measurement or observation at a vertex or a region of the net provides, through Bayesian statistics, a re-evaluation of these probabilities after a measurement.   This is illustrated in Figure 2 where an event at A is more likely to have been caused by an event at $B$ than $C$ and $D$ is an impossibility due to zero connectivity between the paths. Bayes theorem provides a way of translating this common sense concept into a formal probabilistic context since $P(A \mid B)  > P(A \mid C) > P(A \mid D)$.

\begin{figure}[h]
\centering
\begin{minipage}{14pc}
\centering
\includegraphics{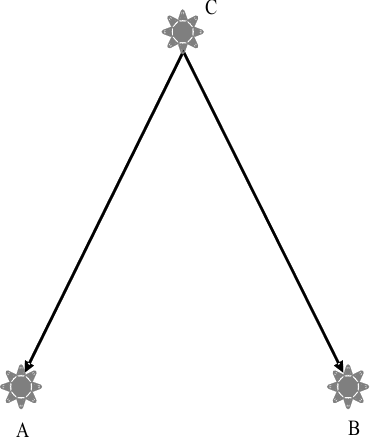}
\caption{\label{label}Reichenbach “conjunctive” fork linking events A and B with common cause C. C is earlier in time than simultaneous events A and B.}
\end{minipage}\hspace{2pc}%
\begin{minipage}{18pc}
\includegraphics{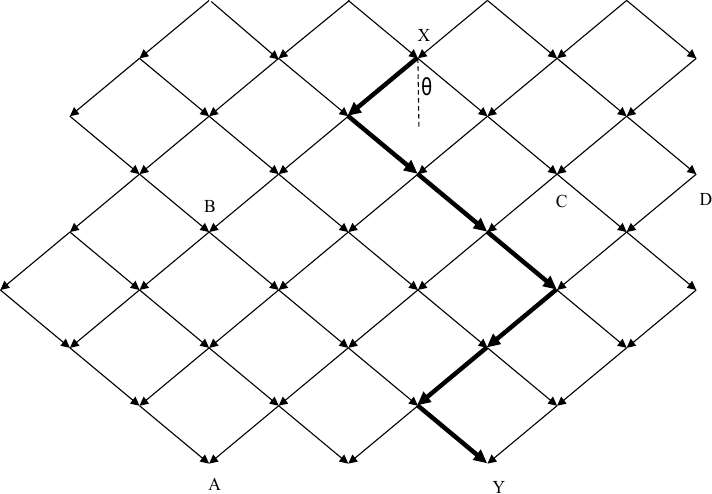}
\caption{\label{label}Causal net showing causal chain from X to Y.}
\end{minipage} 
\end{figure}

\section{Relativistic causal nets for a free particle}
First, we will consider the simple case of a particle randomly diffusing on the causal net shown in Figure 2. In this model time and space can be discretely ``counted'' by attaching an integer to each of the vertex points but there is no underlying continuous space-time. To relate to conventional mechanics we interpolate this set of integers by a set of real number coordinates. Expecting that space and time have different dimensions we need to introduce a constant $c$ with dimensions [space/time]. The net is then made up of elementary triangles labelled with $(\Delta x, c \Delta t, c \Delta \tau)$ as shown in Figure 3. We have not yet added any specific interpretation to these quantities. However, to guarantee invariance of causality on the net we impose $c$ as the speed of light [16]. Since, from geometry,  ${\Delta x / c \Delta t} = \sin \theta \le 1$, we then identify $\Delta x$ and $\Delta t$ as relativistic space-time intervals in an observer frame $S'$ and $\Delta \tau$ as the particle proper time interval in its rest frame $S$. The net geometry guarantees the invariant space-time interval

\begin{equation}
{(c \Delta \tau)^2 = (c \Delta t)^2 - (\Delta x)^2.}
\end{equation}

Having abandoned the concept of absolute and continuous space-time we need to define the observed velocity in terms of finite differences. The definition we shall adopt is  $ v = \Delta x / \Delta t $ which we equate to the expectation of the velocity on the causal net. The two time intervals are then related by $ \Delta \tau = \Delta t / \gamma $ where $\gamma = {1 / \sqrt{1-{v^2 / c^2}}} $ is the Lorentz factor specifying the net angles $\cos \theta = {1 / \gamma}$ and ${\sin \theta = {v / c}}$.

We now specialise to the case of the motion of a free particle. Clearly Eq. (1) and thus the net can be scaled by a factor. If we identify this with the particle rest mass $m$ then from Eq. (1) we then have the relativistic dispersion relation $ E^2=p^2c^2+m^2c^4$ where E is the particle energy $ E = \gamma mc^2 $ and  $ p = \gamma mv $ the momentum. We can further rearrange to derive a second useful invariant relation $-mc^2 \Delta \tau = p \Delta x - E \Delta t$ and a third, the Lagrangian for a free particle $L = -mc^2/\gamma = pv-E = pv-H$ where $H$ is the Hamiltonian.

From our definition of simultaneity and the geometry of the net we can see that the invariant relations provide an action $\sum {p \Delta x}$ which is the same on the lattice for all paths between two events. This is a restatement of Maupertuis principle, which is a weak form of the well known principle of least action, that is the integral $ \int {p \, dx}$ is stationary. On our net if the action $\sum {p \Delta x}$ differed for different trajectories then this would rule some trajectories as physically inadmissible. Therefore we conclude that $\sum {p \Delta x}$ is the same on the lattice for all paths between two events which implies that $ {p \Delta x}$ is a constant $\eta$ for a valid causal net. 

\begin{figure}[h]
\centering
\begin{minipage}{14pc}
\includegraphics{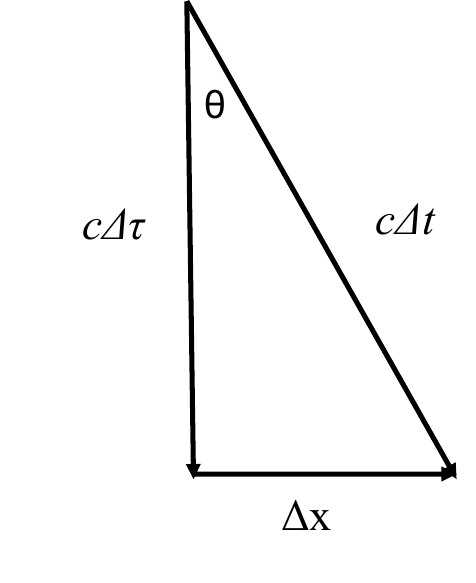}
\caption{\label{label}The elementary space-time “triangle” for the causal net.}
\end{minipage}\hspace{2pc}%
\begin{minipage}{14pc}
\includegraphics{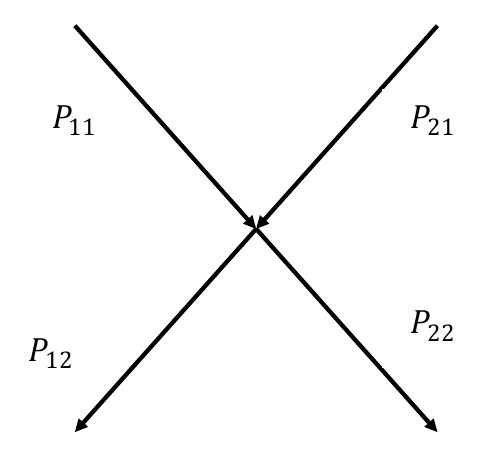}
\\*
\caption{\label{label}A vertex (1,2) on the causal net with associated probabilities.}
\end{minipage} 
\end{figure}

To impose our imperfect correlation of events we shall assume that there is an indeterminism or randomness to the particle motion at each net vertex. We shall make the assumption that this indeterminism is governed by Eq. (1) on the causal net. Thus a particle in its own rest frame $S$ over interval $\Delta \tau$ moving at a speed $\vert {v} \vert$ in inertial frame $S'$ can move to a position $\pm \Delta x$ in time $\Delta t$. This produces a random trajectory in space-time (Fig. 2).

Initially we shall consider ``classical'' or non--quantum probabilities in construction of the causal net. Consider an individual vertex on the net and label the incoming probabilities on row 1 $P_{11}$ and $P_{21}$  and outgoing probabilities on row 2 $P_{12}$  and  $P_{22}$ (Fig. 4).  Probability is conserved at the vertex and the total probability at a vertex is given by ${\Omega}_{12}=P_{11}+P_{21}$. If the average velocity measured on the lattice is uniform then  $P_{11}=P_{22}$ and $P_{12}=P_{21}$. This implies that the probabilities ``cross'' at each vertex without actually interfering although the probabilities are coupled. We shall see that this non-interacting case corresponds to the equilibrium case of a free particle. If we consider normalised branching probabilities at the vertex defined as $\hat{P}_{11}+\hat{P}_{21}=1$ then since expected velocity at the vertex is defined to be $v$ we have

\begin{equation}
E[v] = \gamma {\Delta x \over \Delta t}[\hat{P}_{11} - \hat{P}_{21}] = v.
\end{equation}
The branching probabilities are then given by

\begin{equation}
\hat{P}_{11} = {E+mc^2 \over 2E} \qquad \hat{P}_{21} = {E-mc^2 \over 2E}.
\end{equation}
From this we can see that in the low velocity limit  $\vert v \vert \rightarrow 0$ then $\hat{P}_{11}\rightarrow 1$ and $\hat{P}_{21}\rightarrow 0$  and in the high velocity limit $\vert v \vert \rightarrow c$ then $\hat{P}_{11}\rightarrow \hat{P}_{21}\rightarrow 1/2$. The branching ratio $\Gamma$ can be written as a function of $\gamma$ or the net angle $\theta$.
\begin{equation}
\Gamma = {\hat{P}_{11} \over \hat{P}_{21}} = {{E+mc^2} \over {E-mc^2}} = {{\gamma +1} \over {\gamma-1}} = {{1+\cos \theta} \over {1-\cos \theta}}.
\end{equation}
Using the branching probabilities (Eq. 3) we can write a non trivial matrix equation linking the probabilities

\begin{equation}
\vec{P}=
\left(
\begin{matrix}
\hat{P}_{22} \\
\hat{P}_{12}
\end{matrix} \right) =
\left(
\begin{matrix}
\hat{P}_{11} \\
\hat{P}_{21}
\end{matrix} \right) =
\left(
\begin{matrix}
0&\Gamma \\
{1 / \Gamma}& 0
\end{matrix}
\right)
\left(
\begin{matrix}
\hat{P}_{11} \\
\hat{P}_{21}
\end{matrix} 
\right).
\end{equation}

\section{Relativistic quantum mechanics on the causal net}

We shall now see how the causal net is compatible with the quantum mechanics of the Dirac equation for a free particle. We notice that identifying the net constant $\eta$ with Planck's constant $h$ provides the de Broglie relation $\lambda p = h$ [17] with $\Delta x = \lambda /2$, and a Heisenberg like relation $\Delta p \Delta x \sim h/2$ [18,19]. The discrete nature of the net automatically entails a de Broglie relation and an uncertainty principle. The probability is invariant in the rest frame $S$ of the particle and we can write each probability by combining complex probability amplitudes   $P_{ij}={\phi}_{ij}.{\phi}_{ij}^*$ with

\begin{equation}
\phi_{ij} = \sqrt {P_{ij}} e^{-{imc^2 \tau} / \hbar} = \sqrt {P_{ij}} e^{{i(px-Et)} / \hbar},
\end{equation}
which depends on the proper time $\tau $ at the net vertices and $x=x_{ij}$ and $t=t_{ij}$ are defined at the discrete net vertices. The phase is independent of position $x$ for a particular $\tau$ with equivalent phase thus defining simultaneity on the net and is equivalent to a $U(1)$ global gauge invariance.  We can rewrite Eq. (5) as

\begin{equation}
\Phi = 
\left(
\begin{matrix}
\phi_{22} \\
\phi_{12}
\end{matrix} \right) =
\left(
\begin{matrix}
\phi_{11} \\
\phi_{21}
\end{matrix} \right) =
\left(
\begin{matrix}
0&\sqrt{\Gamma} \\
{1 / \sqrt{\Gamma}}& 0
\end{matrix}
\right)
\left(
\begin{matrix}
\phi_{11} \\
\phi_{21}
\end{matrix} 
\right),
\end{equation}
which can be alternatively expressed for Eq. (7) in terms of a unique transfer matrix $M$

\begin{equation}
\Phi = 
\left(
\begin{matrix}
\phi_{22} \\
\phi_{12}
\end{matrix} \right) =
\left(
\begin{matrix}
\phi_{11} \\
\phi_{21}
\end{matrix} \right) =
M
\left(
\begin{matrix}
\phi_{11} \\
\phi_{21}
\end{matrix} 
\right),
\end{equation}
defined as

\begin{equation}
M = 
\left(
\begin{matrix}
\cos{\theta}&\sin{\theta} \\
\sin{\theta}& -\cos{\theta}
\end{matrix}
\right)=
\left(
\begin{matrix}
{1 / \gamma}& {v/c} \\
{v/c}& -{1 / \gamma}
\end{matrix}
\right)=
\frac {1}{E}
\left(
\begin{matrix}
{mc^2}& {pc} \\
{pc}& -{mc^2}
\end{matrix}
\right)=
\frac {H_{D}}{E}.
\end{equation}
Here we recognise $H_{D}$ as the Dirac Hamiltonian for a free particle [3, 20] with defined momentum $p$. To connect with the complete quantum mechanics we note that Eq. (9) can be put in the conventional form [21] by assuming that space-time is locally differentiable at the vertex, allowing us to use the usual momentum operator $\hat{p}$   to replace the momentum eigenvalues $p$. This assumption of differentiability is satisfied if we consider space-time to be a continuum of discrete causal nets, all infinitesimally displaced in space and proper time. We can write

\begin{equation}
\left(
\begin{matrix}
{mc^2}& {c \hat{p}} \\
{c \hat{p}}& -{mc^2}
\end{matrix}
\right)
\Psi =
E \Psi =
i \hbar \frac {\partial {\Psi}}{\partial {t}},
\end{equation}
where we have replaced the probability amplitudes $\Phi$ with the familiar 2 component Dirac spinor $\Psi$ for the free particle [21]
\begin{equation}
\Psi(x,t) = 
\left(
\begin{matrix}
\psi_{1} \\
\psi_{2}
\end{matrix} 
\right)
=\Phi(x,t) =
A
\left(
\begin{matrix}
1 \\
\frac{pc}{E+mc^2}
\end{matrix} 
\right)
e^{{i(px-Et)} / \hbar},
\end{equation}
and A is an appropriate normalisation constant.

\section{Causal net quantum symmetries, spin and Lorentz invariance}

Note that the unique matrix $M$ above is a unitary, orthogonal matrix which provides an SU(2) group transformation corresponding to an improper rotation -- that is a rotation $R( \theta)$ followed by an inversion $\beta$ so $M=\beta R( \theta)$. The matrix provides the transformations for the probabilities  $\vec{P}= M^2 \vec{P} =I\vec{P}$ and probability amplitudes $\Psi=M \Psi$. Importantly because $M^3 = M$ there exists only two levels of symmetry at the vertex and the causal net provides simultaneously both the probabilities and the underlying probability amplitudes. Since it is an improper rotation the symmetry determines a preferred axis which provides helicity along the axis of movement. If we revisit Eq. (7) and consider both possible positive and negative roots we can see that even and odd solutions with helicity $ \lambda = \pm 1$ and positive energy $\epsilon =+1$ are given by 

\begin{equation}
{\Phi}_
{\substack{\epsilon = +1 \\
\lambda = +1}}
=
\left(
\begin{matrix}
\sqrt{P_{11}} \\
\sqrt{P_{21}}
\end{matrix} 
\right)
\qquad
{\Phi}_
{\substack{\epsilon = +1 \\
\lambda = -1}}
=
\left(
\begin{matrix}
\sqrt{P_{11}} \\
- \sqrt{P_{21}}
\end{matrix} 
\right),
\end{equation}
corresponding to transfer matrices $M_{\epsilon =+1, \lambda= \pm 1}=\beta R( \pm \theta)$. Note that in the above and following discussion we have omitted the normalisation constant and phase factor $e^{-{imc^2 \tau} / \hbar}$ since this cancels in both sides of Eq. (8). Until now we have considered only the positive energy states, but negative energy solutions arise from the negative solution of the relativistic dispersion relation $E=\epsilon \sqrt{p^2 c^2 + m^2 c^4} = \epsilon \vert E \vert $ with $\epsilon = \pm 1$. This results in a reversal of the branching probabilities in Eq. (3) and two additional possible even and odd spinor solutions

\begin{equation}
{\Phi}_
{\substack{\epsilon = -1 \\
\lambda = +1}}
=
\left(
\begin{matrix}
-\sqrt{P_{21}} \\
\sqrt{P_{11}}
\end{matrix} 
\right)
\qquad
{\Phi}_
{\substack{\epsilon = -1 \\
\lambda = -1}}
=
\left(
\begin{matrix}
\sqrt{P_{21}} \\
\sqrt{P_{11}}
\end{matrix} 
\right),
\end{equation}
for transfer matrices $M_{\epsilon =-1, \lambda= \pm 1}= - \beta R( \pm \theta)$. If we include the negative energy states then, by combining all 4 net solutions above (Eq. 12 and Eq. 13), we can write 4 orthogonal 4-vectors which for helicity $\lambda= \pm 1$ and energy $E=\epsilon \vert E \vert$ with $\epsilon= \pm 1$ are
\begin{equation}
\Psi_
{p, \epsilon , \lambda =+1}
=A
\left(
\begin{matrix}
1 \\
0 \\
\frac{cp}{E+mc^2} \\
0
\end{matrix}
\right)
e^{{i(px-Et)} / \hbar}
\qquad
\Psi_
{p, \epsilon , \lambda =-1}
=
A
\left(
\begin{matrix}
0 \\
1 \\
0 \\
\frac{-cp}{E+mc^2} 
\end{matrix}
\right)
e^{{i(px-Et)} / \hbar}.
\end{equation}
Using the 4 possible transfer matrices $M_{\epsilon = \pm 1, \lambda= \pm 1}$ then the 1+1 dimension 4-matrix Dirac equation is
\begin{equation}
\left(
\begin{matrix}
{mc^2}& {c \beta \hat{p}} \\
{c \beta \hat{p}}& -{ mc^2}
\end{matrix}
\right)
\Psi =
E \Psi =
i \hbar \frac {\partial {\Psi}}{\partial {t}}.
\end{equation}
Now this Dirac equation and the spinor wavefunction Eq. (15) correspond to exactly the conventional 3+1 dimension Dirac spinor for the special case of the particle moving along the x-axis and with a well defined spin aligned parallel and antiparallel with the x-axis [21].

The causal net is also consistent with the extraordinarily simple Foldy–-Wouthuysen representation [22] of the Dirac equation where the positive and negative energy states are decoupled through a rotation of $\theta$ (the net angle) of the Dirac Hamiltonian. For example, one Foldy–-Wouthuysen 1+1 dimension state is given by the rotation through $\theta /2$ of the Dirac state Eq. (12) since $cos(\theta /2) = \sqrt{P_{11}}$ and $sin(\theta /2) = \sqrt{P_{21}}$

\begin{equation}
{\Phi}_
{\substack{FW \\
\epsilon = +1 }}
=
{R( \theta /2)}
{\Phi}_
{\substack{Dirac \\
\epsilon = +1 \\
\lambda = +1}}
=
\left(
\begin{matrix}
\sqrt{P_{11}}&\sqrt{P_{21}} \\
-\sqrt{P_{21}}& \sqrt{P_{11}}
\end{matrix}
\right)
\left(
\begin{matrix}
\sqrt{P_{11}} \\
\sqrt{P_{21}}
\end{matrix}
\right)
=
\left(
\begin{matrix}
1 \\
0
\end{matrix}
\right).
\end{equation}
Importantly, in this representation, establishing an exact particle position is impossible (there is only a mean position operator) and a particle is viewed as spread out over a finite region of about a wavelength which is consistent with our causal net picture. 

The causal net can also be constructed by a Lorentz boost of the Foldy–-Wouthuysen states. Setting $tanh  \omega= v/c = sin \theta$ then the Dirac spinor Lorentz operator $\hat{S}_{L} (\omega)$ acts for example as

\begin{equation}
{\Phi}_
{\substack{Dirac \\
\epsilon = +1 \\
\lambda = +1}}
=
\hat{S}_{L}
{\Phi}_
{\substack{FW \\
\epsilon = +1 }}
=
\left(
\begin{matrix}
\cosh\omega /2&-\sinh\omega /2 \\
-\sinh\omega /2& \cosh\omega /2
\end{matrix}
\right)
{\Phi}_
{\substack{FW \\
\epsilon = +1 }}
=
\gamma^{1/2}
\left(
\begin{matrix}
\sqrt{P_{11}}&\sqrt{P_{21}} \\
\sqrt{P_{21}}& \sqrt{P_{11}}
\end{matrix}
\right)
\left(
\begin{matrix}
1 \\
0
\end{matrix}
\right)
.
\end{equation}
Applying successive Lorentz boosts reconstructs a symmetric causal network of connected events in any reference frame so, for example $\Phi '(x',t')= \hat{S}_{L} (\hat{\omega })\Phi(x,t)$, for boost $\hat{\omega}$ with $ \omega ' = \omega + \hat{\omega}$. The Lorentz boost to any frame preserves the conservation and composition of probabilities $P_{11}+P_{21} = \gamma (P_{11}-P_{21})=1$ as Eq. (2) to satisfy simultaneity on the net.

\section{The 3+1 dimension Dirac equation}

To extend to the general 3+1 dimension case we must consider transformations of the causal net that leave it invariant under spatial direction of velocity $\vec{v}$. Using polar coordinates then for momentum $\vec{{p}}= \vert \vec{p} \vert (sin{\vartheta}cos{\varphi},sin{\vartheta}sin{\varphi},cos{\vartheta})$ we can expect that the wavefunction components become dependent on the coordinates $(\vartheta , \varphi)$ so $\sqrt{P_{ij}}$ becomes $\sqrt{P_{ij}} {\chi}(\vartheta , \varphi)$. Following Dirac's convention [3] we can replace the 1 dimension momentum operator ${\hat{p}}$ with the 3 dimensional momentum operator $(\vec{\sigma }.
\vec{p})$, formed from Pauli matrices $ \sigma_{k}$ $(k=1,2,3)$. By definition this momentum operator provides the relation $(\vec{\sigma}. \vec{p}) \chi_{\pm} = \vert \vec{p} \vert \chi_{\pm}$ with two eigenvectors
$
{\chi}_{+}
=
(
\cos{\vartheta/2},
e^{i \varphi} \sin{\vartheta/2}
)
$
,
$
{\chi}_{-}
=
(
-e^{-i\varphi}\sin{\vartheta/2},
\cos{\vartheta/2}
).
$
The general solutions for the wavefunction then become from (Eq. 12 and Eq. 13) four 4-component orthogonal vectors corresponding to up and down spin $S=\pm 1/2$ with positive and negative energies $\epsilon = \pm 1$. Omitting the phase factors and normalisation constant these are
\begin{equation}
{\Psi}_
{\substack{\epsilon = +1 \\
S = +1/2}}
=
\left(
\begin{matrix}
\sqrt{P_{11}} {\chi}_{+} \\
\sqrt{P_{21}} {\chi}_{+}
\end{matrix} 
\right)
\qquad
{\Psi}_
{\substack{\epsilon = +1 \\
S = -1/2}}
=
\left(
\begin{matrix}
\sqrt{P_{11}} {\chi}_{-} \\
- \sqrt{P_{21}} {\chi}_{-}
\end{matrix} 
\right),
\end{equation}

\begin{equation}
{\Psi}_
{\substack{\epsilon = -1 \\
S = +1/2}}
=
\left(
\begin{matrix}
-\sqrt{P_{21}} {\chi}_{+} \\
\sqrt{P_{11}} {\chi}_{+}
\end{matrix} 
\right)
\qquad
{\Psi}_
{\substack{\epsilon = -1 \\
S = -1/2}}
=
\left(
\begin{matrix}
\sqrt{P_{21}} {\chi}_{-} \\
\sqrt{P_{11}} {\chi}_{-}
\end{matrix} 
\right).
\end{equation}
These are the general solutions to the conventional 3+1 dimension Dirac equation

\begin{equation}
\left(
\begin{matrix}
{mc^2}& {c(\vec{\sigma }.\vec{p})} \\
{c(\vec{\sigma }.\vec{p})}& -{mc^2}
\end{matrix}
\right)
\Psi =
E \Psi =
i \hbar \frac {\partial {\Psi}}{\partial {t}}.
\end{equation}
Thus for a 3 dimensional space we require all 3 Pauli matrices to construct the vector $ \vec{\sigma}$ and the dimensionality of space defines the Pauli matrices. A plane wave solution to the Dirac equation has a unique velocity or space direction and the causal net is constructed along this direction in 3 dimensional space.

\section{The effect of a potential on a causal net and the Pauli equation}

The case we have examined is that of a free particle but we could include a potential $V$ on the causal net since $E$ can be replaced by $E - V$ in the construction of the lattice and the branching ratios.  Returning to the 1+1 dimension case, between two media with different scalar potentials the net is compressed or stretched in space in the potential region with a form similar to Snell's law $cos \theta_{2} / cos \theta_{1} = E/(E-V)$. We can write Eq. (10) conveniently as

\begin{equation}
M =
\left(
\begin{matrix}
\cos{{\theta}_{2}}& \sin{{\theta}_{2}} \\
\sin{{\theta}_{2}}& -\cos{{\theta}_{2}}
\end{matrix}
\right)
=
\frac {1}{E-V}
\left(
\begin{matrix}
{mc^2}& {pc} \\
{pc}& -{mc^2}
\end{matrix}
\right)=
\frac {H_{D}}{E-V}.
\end{equation}
If we consider invariance under a local gauge transformation $U$ then in general $M(U \Phi) \ne (U \Phi)$ so to retain invariance we must add an additional term to the causal net Dirac equation or Lagrangian which corresponds to a gauge potential term. The simple conjunctive fork (Fig. 1) is effectively replaced with an ``interactive'' fork [23], representing the interaction between different causal nets. To illustrate, consider the special case of a transformation where the proper time interval $\Delta \tau$ is unchanged by a potential. The triangle in Figure 3 is deformed by an amount $\delta \tau$ in time and $\delta x$ in space
${(c \Delta \tau)}^2 = {(c (\Delta t-\delta t))}^2 - {(\Delta x - \delta x)}^2$. If we write $eA_{0} = \gamma mc \delta t / \Delta t$ and $eA_{1}=\gamma mc \delta x / \Delta t$ then we have the dispersion relation for an electron of charge $e$ in an electromagnetic field $(A_{0},A_{1})$
as $
{(E-e{A}_{0})}^2 = {(pc-e{A}_{1})}^2 +m^2 c^4
$
and the corresponding transfer matrix $M$ is given by

\begin{equation}
M =
\left(
\begin{matrix}
\cos{{\theta}_{2}}& \sin{{\theta}_{2}} \\
\sin{{\theta}_{2}}& -\cos{{\theta}_{2}}
\end{matrix}
\right)
=
\frac {1}{E-e{A}_{0}}
\left(
\begin{matrix}
{mc^2}& {pc-e{A}_{1}} \\
{pc-e{A}_{1}}& -{mc^2}
\end{matrix}
\right).
\end{equation}
If as Eq. (20) we embed the causal net in a continuous  3 dimensional space we can replace p with the 3 dimensional momentum operator and can consider the non-relativistic case of motion in a weak field. If we neglect the smaller component of the spinor and $E'=E-mc^2$ we have, following [20], the Pauli equation for a non-relativistic spin-1/2 particle.
\begin{equation}
\left(
\frac{(\vec{p}-e\vec{A})^2}{2m}
+eA_{0}
-\frac{e \hbar}{2mc}
(\vec{\sigma}.\vec{H})
\right)
\psi
=E' \psi.
\end{equation}

\begin{figure}
\centering
\includegraphics{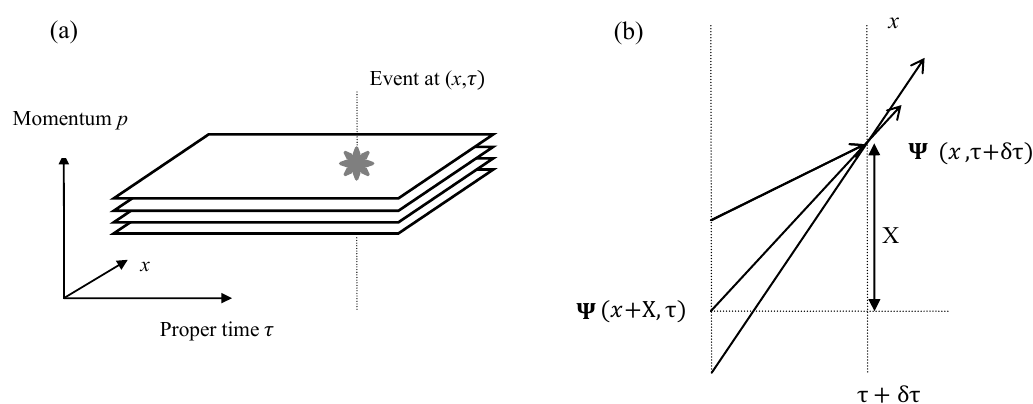}
\caption{(a) A stacked ``deck'' of causal nets for different momentum states and (b) causally connected paths traversing a single space-time event.}
\end{figure}

\section{Momentum states and the Feynman path integral}

It is interesting to consider the more general case of a range of momentum states with each momentum state occupying an individual causal net. This can be visualised in Figure 5(a) as a stacked ``deck'' of infinitely extended causal nets. Consider an event in  at $(x,\tau+\delta \tau)$ and the prior events that are causally connected from a earlier slice of proper time at $\tau$, which are given by different space points $x$ from each momentum net. If we sum the different spinor components contributing to the overall probability amplitude at $(x,\tau +\delta \tau)$ and include the change in phase over interval $\delta \tau$ from Eq. (6) we have
\begin{equation}
\Psi (x,\tau + \delta \tau) 
=
\sum_{\substack{causally \\
connected \\
points}}
\Psi (x+X,\tau)
e^{-{imc^2 \delta \tau / \hbar}},
\end{equation}
where $X$ is the relative space coordinate (Fig. 5b). For one casual net representing a free particle with a single momentum state Eq. (24) is trivial since velocity and probability are uniform across the net with only the phase varying with $\tau$ so
$
\Psi (x,\tau + \delta \tau) 
=
\Psi (x+X,\tau )
e^{-{imc^2 \delta \tau / \hbar}}
$
providing a simple delta function propagator for proper time interval $\delta \tau$,
$
K (x,x+X;\tau + \delta \tau) 
=
\delta (X)
e^{-{imc^2 \delta \tau / \hbar}}
$.
However, if there is a continuum of momentum nets by geometry the sum in Eq. (24) selects a single probability amplitude contribution from each net with momentum $p=mX/\delta t$ for a given relative position $X$. Writing the relativistic infinitesimal action $S_{rel}(\delta \tau) = -mc^2 \delta \tau$ we can write Eq. (24) as an integration 

\begin{equation}
\Psi (x,\tau + \delta \tau) 
=
\int\limits_{-\infty}^{\infty}
{\Psi}_{p} (x+X,\tau) 
e^{{iS_{rel}(\delta \tau)  / \hbar}}
\mathrm{d}X,
\end{equation}
where ${\Psi}_{p}$  denotes the spinor with momentum $p=mX/\delta \tau$. If we consider the non-relativistic limit where one spinor component $\psi$  dominates and $\tau \rightarrow t$ we can use the semi-classical action 
${S}_{cl}
=
\int
(m/2)
{({\mathrm{d}x}/{\mathrm{d}t})}^2
\mathrm{d}t
$
 and write this infinitesimal path integral in the limit of large time interval $T$ to give the conventional Feynman path integral [4]

\begin{equation}
\psi (x, t+T) 
=
\int\limits_{-\infty}^{\infty}
{K}_{0} (x,X;T) \psi(X,t)
\mathrm{d}X,
\end{equation}
where  
$
{K}_{0} (x,X;T)
=
\sqrt {(m/ 2 \pi i \hbar T)}
e^{i m X^2  / 2 \hbar T}
$
is the free particle propagator for the Schr\"{o}dinger equation. The causal net model is thus consistent with the quantum mechanical summation of paths and solutions to various problems such as slit diffraction using Feynman integrals [4].

\section{Non-Euclidean space-time, general relativity and mass}

In non-Euclidean curved space-time our elementary triangles (Fig. 3) comprising our causal net will become distorted and we can no longer apply Pythagoras' theorem to evaluate the space-time interval. In the language of general relativity the space-time interval is given by the metric $g_{{\mu \nu}}$ so $ds^2=g_{{\mu \nu}}dx^\mu dx^\nu$. Previously, we have considered the special case of the Minkowski metric ${\eta}_{\mu \nu}$ for flat space-time but general relativity considers Riemann spaces that have quadratic metric equations and are characterised as locally flat. Considering a small displacement in space $\Delta x$ from a point $x$ using Taylor expansion we have the metric
$
{g}_{\mu \nu}(x+ \Delta x)
={\eta}_{\mu \nu}
+
\frac{1}{2}
{g}_{\mu \nu , \rho \sigma}
\Delta {x}^\rho
\Delta {x}^\sigma.
$
We might assume that variation to the scalar Minkowski action $S_{rel}$ would produce a correction ${S_{g}}$ which, to be a scalar under general coordinate transformations, can only include second order derivatives of the metric. Mathematically, the simplest curvature scalar is the Ricci scalar $R = g^{\mu \nu}R_{\mu \nu}$ formed from the Ricci curvature tensor $R_{\mu \nu}$. We can postulate that the simplest additional action might be of the form
${S}_{g}
=
B
\int
R
\mathrm{d^4}x
$. 
If we arbitrarily set the constant $ B = -{\varepsilon}_{0}/{16 \pi G c^4} $ where $ {\varepsilon}_{0} $ is a ``density'' and $G$ Newton's gravitational constant then we have Einstein's \emph{unimodular} gravity. If we further impose the density as $ {\varepsilon}_{0} = -\sqrt {-g} $ where $ g= det({g}_{\mu \nu}) $ then we recover the Einstein--Hilbert action

\begin{equation}
{S}_{g}
=
- \frac{1}{16 \pi G c^4}
\int
R
\sqrt {-g}
\mathrm{d^4} x.
\end{equation}
Thus the causal net model, although a microscopic theory, would appear to be consistent with the macroscopic theory of general relativity if the elementary triangles of our causal net model are distorted to ``tile'' curved space--time between causally connected possible events to preserve our definition of simultaneity. 

Finally, in our model the particle mass $m$ is an arbitrary scaling constant but if we assume a 1+1 dimension Laplace equation describes the diffusive causal paths on the net, then by imposing a Dirichlet boundary condition, the uncharged mass term of the kth fermion follows an exponential spectrum with $m_{k}=exp(ak+b)$ where $a$ and $b$ are constants. 

\section{Discussion}
By considering simple casual connections between elementary events, based on Reichenbach's principle of common cause, we have constructed a causal model where the Dirac equation and the fermion particles it describes are seemingly ``emergent'' properties. The simplest causal network describes exactly the Dirac equation and provides quantum mechanical statistics and major quantum phenomena (diffraction, wave-particle duality, uncertainty ...). In an inertial frame an observer will view an ordered series of events in space--time as an entity behaving as either a wave or a particle, depending on how the experimental measuring setup is conceived, and thus exhibits wave-particle duality exactly as proposed by de Broglie. Geometrical quantities of the causal net correspond to measurable physical qualities: mass (scaling factor), momentum and energy (net angle and geometry) and potentials and forces (change of net angle). The global gauge symmetry of the net provides quantum phase and the other degenerate solutions arising from the symmetries of the net are equivalent to the Dirac spin and negative energy states. The discretisation of the net infers similarities to the Heisenberg uncertainty principle and a ``stack'' of nets provides, consistent with the Feynman path integral approach, quantum phenomenon such as diffraction. Distorting the causal net in non-Euclidean space-time suggests analogies with general relativity. Also imposing extra internal Lie symmetries corresponds to the case of ``interacting'' causal forks and leads to the introduction of gauge forces such as electromagnetism. Conventional continuous physics, with its well known differential equations, can perhaps be viewed as being ``emergent'' from a discrete underlying causal net. None of these emergent aspects of our causal net would have been apparent from our simple starting point of an equilibrium distribution of ordered events. Fermion particles can be considered as quasiparticles of the causal network composed of possible and actual events analogous to holes, phonons and recent magnetic monopoles in spin-ice. The quantum measurement ``problem'' is reduced to simple Bayesian statistics and there is no wavefunction ``collapse''. The quantum to classical transition becomes a straightforward  feature of the resolution of the net and conventional quantum mechanics can be tentatively viewed as nature's causal ``equilibrium'' condition. Lastly, the model is compatible with both subjective (Kant, Hume) and objective (Reichenbach) philosophical models of causality.

\end{document}